\documentclass[
 reprint,showkeys,showpacs,preprintnumbers,
 amsmath,amssymb,
 aps,
 prl,
 longbibliography,
 lengthcheck,%
]{revtex4-1}

\usepackage{graphicx}
\usepackage{bm}
\usepackage{amsmath}
\usepackage[bookmarks=false]{hyperref}
\usepackage{dcolumn}

\begin{document}

\title{Implementing the Deutsch's algorithm with spin-orbital angular momentum of photon without interferometer}
\author{Pei Zhang,$^{1 *}$ Yan Jiang,$^{1,2} $ Rui-Feng Liu,$^{1} $ Hong Gao,$^{1}$ Hong-Rong Li,$^{1}$ and Fu-Li Li$^{1} $}
\address{$^{1}$MOE Key Laboratory for Nonequilibrium Synthesis and Modulation of Condensed Matter, Department of Applied Physics, Xi'an Jiaotong University, Xi'an 710049, China\\
$^{2}$Hefei National Laboratory for Physical Sciences at Microscale and Department of Modern Physics, University of Science and Technology of China, Hefei 230026, China\\
*Email address: zhangpei@mail.ustc.edu.cn}

\begin{abstract}
Deutsch's algorithm is the simplest quantum algorithm which shows the acceleration of quantum computer. In this paper, we theoretically advance a scheme to implement quantum Deutsch's algorithm in spin-orbital angular momentum space. Our scheme exploits a newly developed optical device called "q-plate", which can couple and manipulate the spin-orbital angular momentum simultaneously. This experimental setup is of high stability and efficiency theoretically for there is no interferometer in it.
\end{abstract}

\maketitle

\section{Introduction}

In recent few decades, quantum information science has made a great development in quantum computation \cite{Nielsen}, quantum cryptography \cite{cryptography}, quantum metrology \cite{metrology}, quantum lithography \cite{lithography} \emph{et al}. Maybe the most profound application is quantum computer, which promises exponentially faster operation for particular tasks, like factoring large integer \cite{integer} and Grover's algorithm for accelerating combinatorial searches \cite{Grover}. So the problem of how to implement a quantum computation process in a realistic physical system comes out. Early proposals rely on nonlinear couplings between different optical modes, but achieving such couplings at sufficient strengths are difficult in technical \cite{coupling}. In 2001, E. Knill \emph{et al} demonstrated that a system with linear optics requires single photon sources, beam splitters, phase shifters, photon-detectors, and feedback from photon-detector outputs is sufficient for efficient quantum computation (QC) \cite{knill}. This stimulates the researchers paying more attentions to realize QC on optical system\cite{brien,gasparoni}. Quantum algorithm is the soul of QC. The first quantum algorithm was proposed by Deutsch in 1985 \cite{DA1}, and then improved by Deutsch and Jozsa in 1992 \cite{DA2}.
This fundamental quantum algorithm has been theoretically studied and experimentally realized in many kinds of system containing the optical system \cite{Oliveira05,tame,vallone,pzhang}.

As we know, photon carries spin angular momentum (SAM) and orbital angular momentum (OAM). The spin part is associated with the circular polarization of light. For each photon, the SAM is $\sigma\hbar$ ($\sigma=+1$ for left circularly polarized light and $\sigma=-1$ for right circularly polarized light). The OAM is associated with the azimuthal phase of the light field. Allen \emph{et al} \cite{Allen1}, showed that any photon with a phase dependence of the form $exp(\emph{il}\varphi)$, carries an OAM of $\emph{l}\hbar$, and $\emph{l}=0,\pm1,\pm2,\cdots$. So the OAM can be defined as an orthogonal infinite-dimensional discrete Hilbert space, offering a promising resource for high-dimensional quantum information protocols \cite{Allen2,qudit}. As most commercial lasers operate in their fundamental transverse mode producing a Gaussion output beam, several proposals have been developed to generate and manipulate high order laser modes and OAM. Such as, holograms \cite{hologram}, spiral phase plates \cite{phaseplate} and cylindrical lens mode converters \cite{lens}. Recently, another new optical device, called q-plate \cite{qplate1,qplate2,qplate3}, can manipulate the coupling between SAM and OAM, and has been used to realize the quantum information process \cite{Nagali09a,Nagali09b,Slussarenko10}.
Single-photon few-qubit system is one of the widely used optical systems to build the deterministic quantum information processor (QIP). And photon's SAM and OAM are very good candidates to realize the single-photon few-qubit system. In the previous works, A. N. de. Oliveira \emph{et al.} have experimentally tested the Deutsch's algorithm by using single-photon two-qubit (SPTQ) system \cite{Oliveira05}. They employed the light's polarization and Hermit-Gaussian modes as control and target qubit for controlled-NOT gate (CNOT), respectively. However, they implemented the Deutsch's algorithm with a Mach-Zehnder interferometer, which was not stable enough except additional equipment (such as feedback controller) to keep the relative phase between the two arms fixed. We have also implemented Deutsch's algorithm with a Sagnac interferometer \cite{pzhang}. It's a robust setup with high stability and the results fit the theory very well, but the setup is still a little complex. In this paper, we advance a proposal to implement the Deutsch algorithm by using photon's SAM and OAM. Benefiting to the q-plate, our scheme of Deutsch's algorithm is without any interferometer, which makes it very simple and with high stability and efficiency.

\section{Deutsch's algorithm}
Deutsch's algorithm is a solution to discover the state of a Boolean function $f(x)$ to be constant $(f(0)=f(1))$ or balanced $(f(0) \neq f(1))$, and can be described as follows. Suppose we are given a boolean function $f(x)$, the input $x$ is 0 or 1, the output $f(x)$ is 0 or 1, too. We can sort four possible results into two classes. One is called constant function, which contains $f(x)=0$ and $f(x)=1$ or $f(0)=f(1)$; the other is called balanced function, which contains $f(x)=x$ and $f(x)=x\oplus1$ or $f(0)\neq f(1)$ ('$\oplus$' represents Boolean addition). In classical computation, we have to evaluate the function twice to discover the class of $f(x)$. Because of the quantum parallelism, we can send a superposition state of well defined basis. So quantum computation can discover the class of $f(x)$ by evaluating the function for just one time.
Fig. 1 is the quantum circuit implementing Deutsch's algorithm \cite{Nielsen}. $H$ is Hadamard operator which makes $|0\rangle$ and $|1\rangle$ into $(|0\rangle+|1\rangle)/\sqrt{2}$ and $(|0\rangle-|1\rangle)/\sqrt{2}$, respectively. And operator $U_{f}$ makes the input qubits $|x\rangle$ and $|y\rangle$ into $|x\rangle$ and $|y\oplus f(x)\rangle$, respectively.

\begin{figure}[tbh]
\includegraphics[width=6cm]{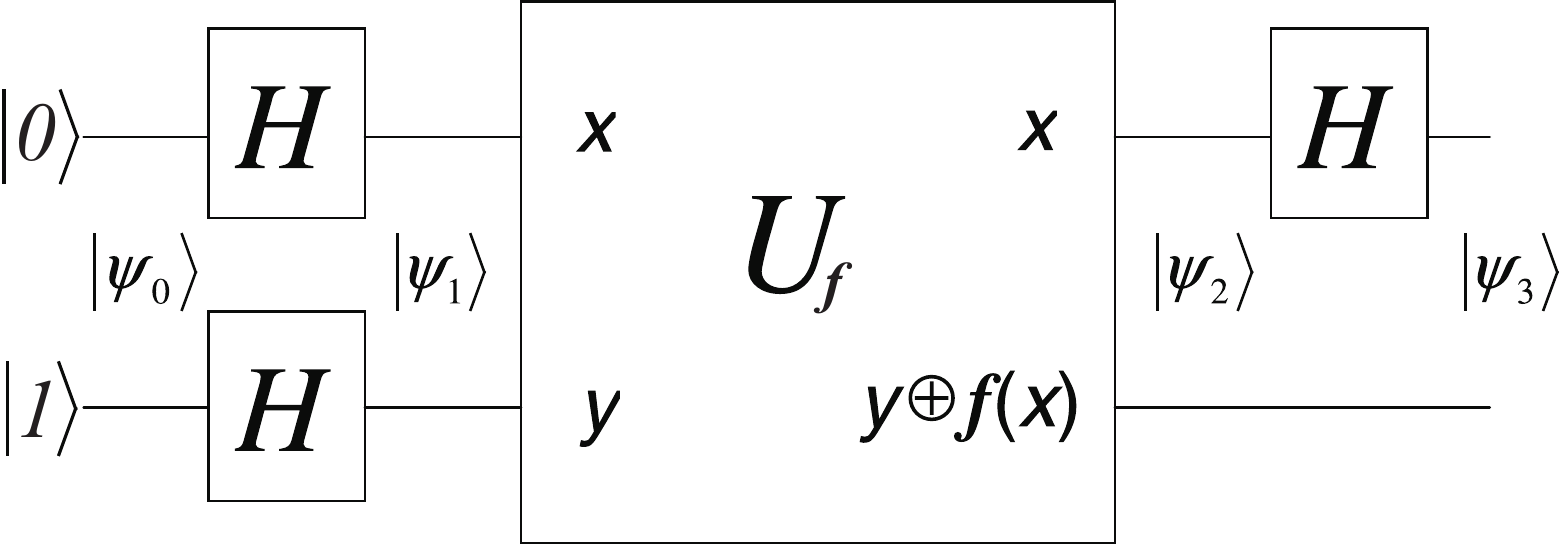}
\caption{Quantum circuit of Deutsch's algorithm. $H$ is the Hadamard gate, and $U_{f}$ takes the input $|x, y\rangle$ into $|x, y\oplus f(x) \rangle$. }
\end{figure}

Traditionally, the input states for the first qubit is $|0\rangle$ and the second qubit is $|1\rangle$, that is $\left\vert \Psi _{0}\right\rangle = \left\vert 0\right\rangle \left\vert1\right\rangle$. So after the Hadamard operation:
\begin{equation}
\left\vert \Psi _{1}\right\rangle =H\left\vert \Psi _{0}\right\rangle=(\left\vert 0\right\rangle
+\left\vert 1\right\rangle )(\left\vert 0\right\rangle -\left\vert1\right\rangle )/2.
\end{equation}%
The output state of $U_{f}$ is:
\begin{equation}
\makeatletter
\let\@@@alph\@alph
\def\@alph#1{\ifcase#1\or \or $'$\or $''$\fi}\makeatother
\left\vert \Psi _{2}\right\rangle=
\begin{cases}
\pm \left[ \frac{\left\vert
0\right\rangle +\left\vert 1\right\rangle }{\sqrt{2}}\right] \left[ \frac{%
\left\vert 0\right\rangle -\left\vert 1\right\rangle }{\sqrt{2}}\right],& f(0)=f(1), \\
\pm \left[ \frac{\left\vert 0\right\rangle -\left\vert
1\right\rangle }{\sqrt{2}}\right] \left[ \frac{\left\vert
0\right\rangle -\left\vert 1\right\rangle }{\sqrt{2}}\right],
&f(0)\neq f(1).
\end{cases}
\makeatletter\let\@alph\@@@alph\makeatother
\end{equation}
After the finally Hadamard gate acting on the first qubit, we can discriminate the function by measuring the state of the first qubit only once: if it is equal to $|0\rangle$ , $f(x)$ is constant, and vice versa.

\section{Q-plate and CNOT gate}

Q-plate is a slab of liquid crystal (or uniaxial
birefringent medium) with special structure, and it is essentially a
retardation wave plate whose optical axis is aligned nonhomogeneously in the
transverse plane in order to create a topological charge $q$ in its
orientation. Each photon being converted
from right circular to left circular changes its spin z-component angular
momentum from $-\hbar $ to $+\hbar $, and the orbital z-component angular
momentum changes $-2q\hbar $. When $q=1$, the total variation of angular
momentum is zero, and there is no net transfer of angular momentum
to the q plate. The plate acts only as a medium for the conversion between
spin and orbital angular momentum. If $q\neq 1$, it will exchange an angular
momentum of $\pm 2\hbar (q-1)$ with each photon. Using Dirac marks, suppose
the initial state is $|1_{s},l_{o}\rangle $ or $|-1
_{s},l_{o}\rangle $, which refers to the photon carrying SAM with $\hbar $
or $-\hbar $ and
OAM with $l\hbar $. The action of a tuned q-plate on this
state can be summarized as follows:
\begin{equation}
(|1_{s},l_{o}\rangle ,|-1_{s},l_{o}\rangle )\overset{q-plate%
}{\longrightarrow }(|-1_{s},(l+2q)_{o}\rangle ,|1
_{s},(l-2q)_{o}\rangle )
\end{equation}%
In this way, for example, a photon in an OAM state $l=0$ is transformed into $l=\pm 2q$, depending on its
polarization.

In this paper, we use q-plates with $q=1$ to carry out our proposal. When circular polarized light crosses such q-plate, the OAM quantum number $l$ of light will acquire a change $\Delta\emph{l}=\pm2$ whose sign depends on the input polarization: positive for left-circular and negative for right-circular. So the action of this q-plate on the states $|1_{s},l_{o}\rangle$ and $|-1_{s},l_{o}\rangle$ can be summarized as follows:
\begin{equation}
(|1_{s},l_{o}\rangle,|-1_{s},l_{o}\rangle)\overset{q-plate}{\longrightarrow}(|-1_{s},(l+2)_{o}\rangle,|1_{s},(l-2)_{o}\rangle).
\end{equation}%
So, for example, a left-circular polarized photon with OAM $l=0$ is transformed into a photon with OAM $l= +2$, and its polarization is transformed into right-circular polarization simultaneously. From above talking, we can see that q-plate can be used to generate higher order Laguerre-Guassian beam from Gaussian beam, and it can couple photon's SAM and OAM generating an entanglement state between these two spaces.

\begin{figure}[tbh]
\includegraphics[width=6cm]{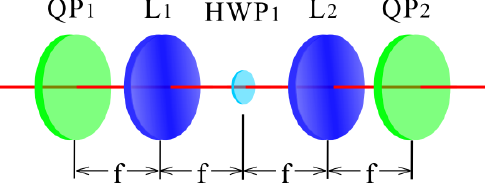}
\caption{The experimental implementation of CNOT gate. The half-wave plate (HWP$_{1}$) placed in the common focal plane of the two lenses (L$_{1}$, L$_{2}$) is inclined at $0^{\circ}$ with vertical direction, and it is used to transform the polarization of photons with OAM $l=0$ only. The focus length of L$_{1}$ and L$_{2}$ is f. The parameter of the q-plates (QP$_{1}$,QP$_{2}$) is $q=1$.}
\end{figure}

In all quantum computation proposals, CNOT gate plays an important role. The essence of CNOT gate is that the value of target qubit is negated if and only if the control qubit has the logical value $"1"$, meanwhile the logical value of the control qubit always keeps unchanged. In the following discussion, we will show how to perform the single-photon two-qubit CNOT gate by using optical devices, where we use SAM as control qubit and OAM as target qubit. The realization of CNOT gate is based on S. Slussarenkothe's work in 2009 \cite{CNOT}, where q-plates were used as essential elements. In this paper, the following logical states are used for all the following quantum logic gates:
\begin{eqnarray}
&&|0,0\rangle=|L,+2\rangle, \ \ |0,1\rangle=|R,-2\rangle  \notag\\
&&|1,0\rangle=|R,+2\rangle, \ \ |1,1\rangle=|L,-2\rangle.
\end{eqnarray}%

Fig. 2 shows the experimental implementation of CNOT gate. The control qubit is in SAM space and the target qubit is in OAM space with the basis $l=\pm2$. The distances between the five devices are same and equal to the focal length of the two lenses. The middle little HWP$_{1}$ only acts on the beam with OAM $l=0$ (transverse distribution is a dot), and leave the OAM $l=\pm4$ (transverse distribution is a ring) components unaffected \cite{CNOT}. The first q-plate is used to take the OAM $l=\pm2$ components of the input beam into OAM $l=0$ and $l=\pm4$, and the second one is used to change them back into OAM $l=\pm2$ with the help of the middle little HWP$_{1}$. Here, we should stress that because of the special optical effect of q-plate, it is convenient to use the special logical states defined in Eq. (5). So the CNOT operation can be written as:
\begin{eqnarray}
&&|0,0\rangle=|L,+2\rangle\overset{CNOT}{\longrightarrow}|L, +2\rangle=|0,0\rangle  \notag\\
&&|0,1\rangle=|R,-2\rangle\overset{CNOT}{\longrightarrow}|R, -2\rangle=|0,1\rangle  \notag\\
&&|1,0\rangle=|R,+2\rangle\overset{CNOT}{\longrightarrow}|L, -2\rangle=|1,1\rangle   \notag\\
&&|1,1\rangle=|L,-2\rangle\overset{CNOT}{\longrightarrow}|R, +2\rangle=|1,0\rangle.
\end{eqnarray}%
We can see this CNOT logic in Eq. (6) is a little different from common one. We can not define logical values without combining SAM and OAM together. However, we can still realize Deutsch's algorithm (discrimination between balanced and constant functions) with above definition.

\section{Scheme for Deutsch's algorithm}

From the preceding description, to physically test the algorithm, we need a device which can implement the $U_f$ operations for the four possible $f(x)$ functions. All the possible functions and operations are summarized in Table I. Corresponding to four different $f(x)$ functions, $U_f$ operations are four two-qubit gates: identity (\emph{I}), NOT, CNOT, and Z-CNOT. Identity gate means nothing changed to target qubit. The operation under our special logical states is:
\begin{eqnarray}
&&|0,0\rangle=|L,+2\rangle\overset{I}{\longrightarrow}|L, +2\rangle=|0,0\rangle  \notag\\
&&|0,1\rangle=|R,-2\rangle\overset{I}{\longrightarrow}|R, -2\rangle=|0,1\rangle  \notag\\
&&|1,0\rangle=|R,+2\rangle\overset{I}{\longrightarrow}|R, +2\rangle=|1,0\rangle   \notag\\
&&|1,1\rangle=|L,-2\rangle\overset{I}{\longrightarrow}|L, -2\rangle=|1,1\rangle.
\end{eqnarray}%
NOT gate means the target qubit always flips no matter which state the control qubit is. It can be wrote as these:
\begin{eqnarray}
&&|0,0\rangle=|L,+2\rangle\overset{NOT}{\longrightarrow}|R, -2\rangle=|0,1\rangle  \notag\\
&&|0,1\rangle=|R,-2\rangle\overset{NOT}{\longrightarrow}|L, +2\rangle=|0,0\rangle  \notag\\
&&|1,0\rangle=|R,+2\rangle\overset{NOT}{\longrightarrow}|L, -2\rangle=|1,1\rangle   \notag\\
&&|1,1\rangle=|L,-2\rangle\overset{NOT}{\longrightarrow}|R, +2\rangle=|1,0\rangle.
\end{eqnarray}%
Z-CNOT gate is short for zero-controlled NOT gate, which means the target qubit flips when control qubit is logical value '0' compared with CNOT gate which the target qubit flips when control qubit is logical value '1'. The Z-CNOT logic can be written as below:
\begin{eqnarray}
&&|0,0\rangle=|L,+2\rangle\overset{Z-CNOT}{\longrightarrow}|R, -2\rangle=|0,1\rangle  \notag\\
&&|0,1\rangle=|R,-2\rangle\overset{Z-CNOT}{\longrightarrow}|L, +2\rangle=|0,0\rangle  \notag\\
&&|1,0\rangle=|R,+2\rangle\overset{Z-CNOT}{\longrightarrow}|R, +2\rangle=|1,0\rangle   \notag\\
&&|1,1\rangle=|L,-2\rangle\overset{Z-CNOT}{\longrightarrow}|L, -2\rangle=|1,1\rangle.
\end{eqnarray}%
So our purpose is to realize four different $U_f$ operations and discriminate them to be balanced or constant on just one running.
\begin{table}
\caption{\label{tab:journals}Four cases of $U_f$ for Deutsch's algorithm}
\begin{ruledtabular}
\begin{tabular}{llll}
\textbf{Class}    &\textbf{Function} &\textbf{Operation} &$\boldsymbol{U_f}$\\
\itshape Constant &$f(x)=0$       &$\left\vert x,y\right\rangle\rightarrow \left\vert x,y\oplus 0\right\rangle$           &Identity (\emph{I})   \\
\itshape Constant  &$f(x)=1$      &$\left\vert x,y\right\rangle\rightarrow \left\vert x,y\oplus 1\right\rangle$           & NOT     \\
\itshape Balanced   &$f(x)=x$      &$\left\vert x,y\right\rangle\rightarrow \left\vert x,y\oplus x\right\rangle$           & CNOT    \\
\itshape Balanced  &$f(x)=x\oplus 1$ &$\left\vert x,y\right\rangle\rightarrow \left\vert x,y\oplus x\oplus 1\right\rangle$  & Z-CNOT         \\
\end{tabular}
\end{ruledtabular}
\end{table}

Fig. 3 is our experimental scheme. It contains four parts: photon source, initial states preparation, operations and detection. The source is a single-photon source which can be obtained by various ways, eg. a deep attenuated coherent light or via a spontaneous parametric down-conversion process. Then we use a polarizer to prepare the initial state of vertical polarization, which can also be recognized as a superposition state of left and right circular polarized state. We can use a special designed computer-generated hologram to prepare the photon's OAM state of $(|2\rangle+|-2\rangle)/\sqrt{2}$. Thus the input state is:
\begin{equation}
|\Psi_{1}\rangle=(\frac{|L\rangle-|R\rangle}{\sqrt{2}})(\frac{|2\rangle+|-2\rangle}{\sqrt{2}}).
\end{equation}%
\begin{figure*}
\includegraphics[width=14cm]{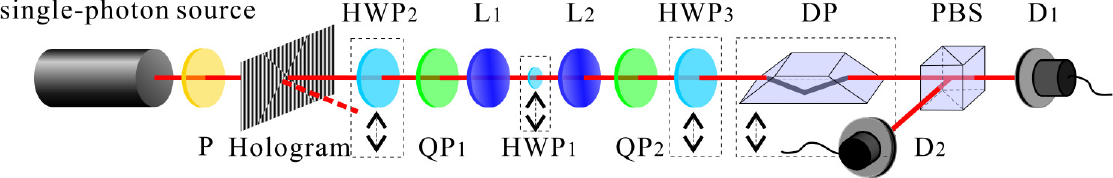}
\caption{\label{fig:wide}Experimental implementation of Deutsch's algorithm. The source is a single photon source. Polarizer (P) and hologram are used to generate the input superposition state $(|L\rangle-|R\rangle)(|2\rangle+|-2\rangle)/2$. Half-wave plates HWP$_{1}$, HWP$_{2}$, HWP$_{3}$ and the Dove prism (DP) are all inclined at $0^{\circ}$, and they can be placed in or moved out from the path to implement the four unitary operations. QP represent q-plate with $q=1$. The middle part is the CNOT gate we have discussed previously. The last polarized beam splitter (PBS) and two detectors (D$_{1}$, D$_{2}$) are used to measure the polarization states of the output photons. }
\end{figure*}
Then the input state will be transformed by the four operations. The operation part is from HWP$_2$ to dove prism (DP).
For the purpose of simply implementing four different operations, two q-plates and two lenses are always in the setup. What we should do is adding or removing HWP(s) and DP, which are all set at $0^{\circ}$, to realize the four different operations.
For the identity operation, nothing happens to the target qubit. We can just leave QP$_{1}$, QP$_{2}$, L$_{1}$, L$_{2}$ in the setup. The polarization is unchanged obviously for missing of HWPs, and the cascaded q-plates keep the OAM unchanged. This acts as an identity operation on photon's SAM and OAM spaces.
For the NOT operation, as Eq. (8) shows, we should change both SAM and OAM. So we can put HWP$_{3}$ and DP in the path based on identity operation, where HWP$_{3}$ flips the SAM and DP set horizontally can reverse the OAM.
We can put HWP$_{1}$ in the path based on identity operation to realize CNOT operation as mentioned previously. The HWP$_{1}$ can only inverse the circular polarization when OAM is $l=0$. Z-CNOT means that the target qubit is reversed when the control qubit is in logical value "0". To carry out this operation, we can put HWP$_{2}$ and HWP$_{3}$ in the path based on CNOT operation. This can be understood as: HWP$_{2}$ flips the control qubit, then photon passes through a CNOT gate, and HWP$_{3}$ flips back the control qubit at last.

When photon passed through the four $U_{f}$ operations, under a simple calculation using the logical states we described previously, Eqs. (6), (7), (8) and (9), the output states $\left\vert \Psi _{2}\right\rangle=U_{f}\left\vert \Psi _{1}\right\rangle$ can be wrote as follow:
\begin{eqnarray}
&&\left\vert \Psi _{2}\right\rangle_{I}=(\frac{|L\rangle-|R\rangle}{\sqrt{2}})(\frac{|2\rangle+|-2\rangle}{\sqrt{2}}); \notag\\
&&\left\vert \Psi _{2}\right\rangle_{NOT}=-(\frac{|L\rangle-|R\rangle}{\sqrt{2}})(\frac{|2\rangle+|-2\rangle}{\sqrt{2}}); \notag\\
&&\left\vert \Psi _{2}\right\rangle_{CNOT}= (\frac{|L\rangle+|R\rangle}{\sqrt{2}})(\frac{|2\rangle-|-2\rangle}{\sqrt{2}}); \notag\\
&&\left\vert \Psi _{2}\right\rangle_{Z-CNOT}= -(\frac{|L\rangle+|R\rangle}{\sqrt{2}})(\frac{|2\rangle-|-2\rangle}{\sqrt{2}}).
\end{eqnarray}
So we can summarize the results as these:
\begin{equation}
\makeatletter
\let\@@@alph\@alph
\def\@alph#1{\ifcase#1\or \or $'$\or $''$\fi}\makeatother
\left\vert \Psi _{2}\right\rangle=
\begin{cases}
\pm \frac{i}{\sqrt{2}}|V\rangle(|2\rangle+|-2\rangle), & f(x)\ is\ constant; \\
\pm \frac{1}{\sqrt{2}}|H\rangle(|2\rangle-|-2\rangle), & f(x)\  is\  balanced.
\end{cases}
\makeatletter\let\@alph\@@@alph\makeatother
\end{equation}
We can clearly see that the output states are orthogonal both in the first qubit and the second qubit for the two different classes of $f(x)$. This is a little different from the original description of Deutsch's algorithm (see Eq. (2)) where the second qubit maintains no changing. This result is caused by the special logical definition differences between mathematical basis and realistic physical condition. So we can discriminate the function to be constant or balanced by detecting the first SAM qubit or the second OAM qubit. As we know, polarization is easy to manipulate, so in the detection part, we use a polarized beam splitter (PBS) to transmit photon with horizontal polarization and reflect vertical polarization. We can get that $f(x)$ is a constant function when $D_{2}$ clicked and $f(x)$ is a balanced function when $D_{1}$ clicked. If using the second qubit to achieve the discrimination, we can place an OAM sorter \cite{sorter1,sorter2} or a simple device such as hologram or spatial light modulator (SLM) instead of PBS.
Anyway, we have achieved our goal to distinguish the two kinds of function (constant and balanced) by one step of calculation.

According the above description, we can implement Deutsch's algorithm using photon's spin-orbital angular momentum.
The main advantages of our scheme are no interferometer in the setup and all photons run in a same optical route, which promise
very high stability and simplicity compared with other protocols. This is mainly benefitted by the q-plate which combines the spin-orbital angular momentum space and cancels
the interferometer. Some other proposals also utilize this property of q-plate to carry out stable quantum information
experiments \cite{Nagali09a,Nagali09b,Slussarenko10,CNOT,qp1,qp2,qp3}. The efficiency of this setup is also very high for
all the operations are deterministic. In the actual experiment, the main lose is in the mode cross-talk on the small HWP$_{1}$. As discussed in Ref. \cite{CNOT}, the waveplate radius must be carefully adapted so to minimize mode cross-talk. The other lose may come from the sequence of two q-plates. However, this lose is under our tolerance. In Ref. \cite{qplate3}, the authors have done an experiment to test the mode generation efficiency of q-plate,
and they found it is almost $97\%$. So in our scheme, the conversion efficiency of the sequence of two q-plates can reach $94\%$ under current technique. Although the OAM state is a little difficult to handle with, we can only detect the polarization of output state. So our scheme is quite simple and stable.

\section{Conclusion}
In a conclusion, we firstly introduce Deutsch's algorithm and an optical element named q-plate, then we use q-plate as the main block to realize Deutsch's algorithm. The Hilbert spaces we use are photon's SAM and OAM. In our design, q-plates and special logical states are used to test Deutsch's algorithm along one optical route without interferometer, which makes our setup more stable and efficient than other experiments. By using the special logical states, we can detect not only the first SAM qubit to realize the discrimination, but also the second OAM qubit. What's more, OAM is a good candidate for encoding high-dimensional quantum state on a single photon \cite{Allen2,qudit,OAM}, so our design is a guidance for realizing more complex and high-dimensional quantum algorithms.

\section{Acknowledgement}

This work is supported by the Fundamental Research Funds for the Central Universities,
Special Prophase Project on the National Basic Research Program of China (Grant No. 2011CB311807),
National Basic Research Program of China (Grant No. 2010CB923102) and National Natural Science
Foundation of China (Grant No. 11004158, 11074198, 11174233 and 11074199).

\end{document}